\documentclass{ptephy}

\usepackage{bm,amssymb}

\graphicspath{{./Figures/}}

\begin{document}
\newcommand{\diff}{\mathrm{d}}
\newcommand{\p}{\partial}
\newcommand{\e}{\varepsilon}
\newcommand{\Diff}{{\mathcal{D}}}
\newcommand{\V}{{\mathcal{V}}}
\newcommand{\up}{\uparrow}
\newcommand{\down}{\downarrow}
\newcommand{\be}{\begin{equation}}      
\newcommand{\ee}{\end{equation}}      
\newcommand{\bea}{\begin{eqnarray}}      
\newcommand{\eea}{\end{eqnarray}}    

\title{Fermionic Functional Renormalization Group Approach to Superfluid Phase Transition}

\author{\name{\fname{Yuya} \surname{Tanizaki}}{1,2,*}, 
\name{\fname{Gergely} \surname{Fej\H{o}s}}{2}, 
and \name{\fname{Tetsuo} \surname{Hatsuda}}{2,3}}

\address{\affil{1}{Department of Physics, The University of Tokyo, Tokyo 113-0033, Japan}
\affil{2}{Theoretical Research Division, Nishina Center, RIKEN, Wako 351-0198, Japan}
\affil{3}{Kavli IPMU (WPI), The University of Tokyo, Chiba 277-8583, Japan}
\email{yuya.tanizaki@riken.jp}}

\begin{abstract}%
A fermionic functional renormalization group (FRG) is applied to describe the superfluid phase transition of the two-component fermionic system with attractive contact interaction. The connection between the fermionic FRG approach and the conventional Bardeen-Cooper-Schrieffer (BCS) theory with Gorkov and Melik-Barkhudarov (GMB) correction are clarified in detail in the weak coupling region by using the renormalization group flow of the fermionic four-point vertex with particle-particle and particle-hole scattering contributions.
 To go beyond the BCS+GMB theory,  coupled FRG flow equations of the fermion self-energy and the four-point vertex are studied 
 under an Ansatz concerning their frequency/momentum dependence.  We found that
 the fermion self-energy turns out to be substantial even in the weak couping region, and the frequency dependence of the four-point 
 vertex is essential to obtain the correct asymptotic-ultraviolet behavior of the flow for the self-energy.
 The superfluid transition temperature and the associated chemical potential are calculated 
 in the region of negative scattering lengths.
\end{abstract}

\subjectindex{A40, A63, B32, I22}

\maketitle
\section{Introduction\label{intro}}

 Superfluidity in  many-fermion systems is one of the central problems in condensed matter, atomic, nuclear and particle physics. Examples include liquid superfluid $^3$He, electron superconductivity, cold atoms, nucleon superfluidity, color superconductivity \cite{Cooper_Feldman201006}, and so on. 
 Among others, ultracold atomic gases play an important role in revealing the nature of superfluidity from weak to strong couplings in a single physical system: inter atomic interactions in dilute systems are rather simple and the strength of the interaction is tunable via the Feshbach resonance technique \cite{arXiv:1306.5785}. 
The experimental discovery of the Bardeen-Cooper-Schrieffer (BCS) to Bose-Einstein condensate (BEC) crossover in two-component fermionic atoms \cite{greiner2003emergence,PhysRevLett.91.250401,PhysRevLett.92.040403,PhysRevLett.92.120403} is a characteristic example of such a high level of control. 

Two seemingly different kinds of superfluidity, BCS superfluid of weakly coupled fermions and BEC of Bose gas, turned out to be the same phenomenon connected via a smooth crossover for two-component fermionic systems. 
On the BCS side, weak attraction causes pairing instability against the Fermi surface leading to the formation of Cooper pairs \cite{PhysRev.108.1175}, while on the BEC side, pairs of fermions form tightly-bound composite bosons and superfluidity occurs due to their condensation at low temperature.
  Furthermore, the crossover between those two forms of superfluidity can be studied by combining the number equation and the assumption of the ground state being a BCS pairing wave function \cite{PhysRev.186.456,leggett1980diatomic,nozieres1985bose}.  

    From a theoretical point of view, quantitative description of fermionic superfluidity in the weak coupling limit requires not only the BCS theory but also its Gorkov and Melik-Barkhudarov (GMB) correction \cite{gorkov1961contribution,PhysRevLett.85.2418}. 
  However, the BCS+GMB theory still ignores higher-order many-body correlations which become important when the scattering length between fermions ($a_s$) becomes large in the so-called unitary regime. 
This is the reason why various non-perturbative techniques such as Monte Carlo simulations \cite{PhysRevLett.96.160402,PhysRevLett.101.090402,PhysRevLett.103.210403,PhysRevA.82.053621,PhysRevB.76.165116}, $\epsilon$-expansion \cite{PhysRevA.75.063618},  the functional renormalization group (FRG) method with auxiliary bosonic field  \cite{birse2005pairing,PhysRevB.78.174528,ANDP:ANDP201010458,PhysRevA.81.063619,arXiv:1010.2890}, 
and the $t$-matrix approach \cite{haussmann1993crossover,PhysRevB.49.12975,PhysRevA.75.023610,arXiv:1109.2307}  have been developed  to attack the problems at and around unitarity.

{  The purpose of this paper  is to develop a fermionic FRG (f-FRG) method without introducing the auxiliary bosonic  field, 
 not only to make a firm connection between the non-perturbative FRG approach and the conventional BCS+GMB theory but also
 to go beyond the BCS+GMB theory on a solid ground}.
 First we will show how BCS+GMB theory is obtained from the renormalization group (RG) flow of the fermionic four-point vertex with particle-particle and  particle-hole interactions, and then we explore the role of the RG flow of the fermion self-energy to go beyond BCS+GMB theory.
We note that such analyses can be best achieved by fermionic FRG without introducing bosonic auxiliary fields. 
Compared with the auxiliary field method, which contains ambiguities in how to introduce the auxiliary field and usually requires \textit{a priori} knowledge on the ground state property of the system, fermionic FRG can provide systematic and unbiased study of interacting fermions \cite{shankar1994renormalization,Salmhofer:2001tr,RevModPhys.84.299,PhysRevB.87.174523}. 

 Throughout this paper, the main focus will be on the weak and intermediate coupling regime, where $a_s$ is negative.  We will clarify the physical meaning of each approximation of f-FRG, and give physical interpretations of our results, based on the analysis of the flow equations in detail. 
Although our formalism itself is not limited to this regime, the case of $a_s >0$  is not accessible at the level of approximations in the present paper, as will be discussed later.  Nevertheless, we will extrapolate our results of the critical temperature and associated chemical potential to the unitary regime to see their qualitative behavior.   

 The contents of this paper is as follows.
 In Sec. \ref{form}, we introduce fermionic FRG formalism and explain how the critical temperature of the superfluid phase transition and the number density of fermions is calculated. 
In Sec. \ref{sec:Flow_Eq}, we construct approximate flow equations of the effective coupling, which reproduce the results of the critical temperature of particle-particle random phase approximation (i.e. BCS theory) and the GMB correction. In Sec. \ref{sec:self_energy}, we concentrate on the flow equation of the self-energy correction, describing its problems and their possible resolution. In Sec. \ref{sec:results}, we solve the coupled flow equations numerically and derive the self-energy correction, the critical temperature, and the associated chemical potential. Sec.\ref{sec:conclusions} is devoted to summary and concluding remarks. The reader can find some useful analytic formulas in the appendices.

\section{Fermionic FRG Formalism\label{form}}
 We consider non-relativistic two-component fermions
 $\psi= \left(
  \begin{array}{c}
  \psi_{\up}     \\
  \psi_{\down}      \\
  \end{array}
\right)$ with a contact interaction:
\begin{equation}
S[\overline{\psi},\psi]=\int_0^{\beta}\diff \tau\int \diff^3\bm{x}
\left[\overline{\psi}\left(\p_{\tau}-{\nabla^2\over 2m}-\mu\right)\psi
+g\overline{\psi}_{\up}\overline{\psi}_{\down}\psi_{\down}\psi_{\up}\right],
\label{intro02}
\end{equation}
where $\beta(=1/T)$, $\mu$, $m$ and $g$ are the inverse temperature, the chemical potential,
 the mass and the bare coupling constant, respectively.
The action $S$ can be written in momentum space as 
\bea
S[\overline{\psi},\psi]&=&\int_p^{(T)} \overline{\psi}(p) G^{-1}(p) \psi(p)\nonumber\\
&+&g\int_p^{(T)} e^{-ip^0 0^+}\int_{q,q'}^{(T)}
\overline{\psi}_{\up}(p/2+q)\overline{\psi}_{\down}({p}/2-q)
\psi_{\down}({p}/2-q')\psi_{\up}(p/2+q'),
\label{form00}
\eea
where $G^{-1}(p)=ip^0+{\bm{p}^2\over 2m}-\mu$ is the inverse propagator
 with $p=(p^0,\bm{p})$. Also, we adopt an abbreviated notation,  
  $\int_p^{(T)}\equiv \int{\diff^3\bm{p}\over (2\pi)^3}T\sum_{p^0}$. 
The factor $\exp(-ip^00^+)$ originates from the  normal ordering of fermionic 
operators in the interaction term.
The classical action (\ref{form00}) is symmetric under global $U(1)$ 
phase rotation, $SU(2)$ spin rotation, and space-time translation.

Following the idea of 
 the FRG method \cite{wetterich1993exact,Morris1,ellwanger1994flow}, we define a scale-dependent generating functional $W_k[\eta,\overline{\eta}]$ by 
\bea
\exp(W_k[\eta,\overline{\eta}])&=&\int \Diff \overline{\psi}\Diff \psi \exp\left[-\Big(S[\overline{\psi},\psi]\right.+\int^{(T)}_p\overline{\psi}(p) R_k(p) \psi(p)\Big)\nonumber\\
&&\left.+\int^{(T)}_p\Big(\eta(p) \overline{\psi}(p)+\overline{\eta}(p)\psi(p)\Big)\right], \label{form03}
\eea
where $\eta$ and $\overline{\eta}$ are fermionic Grassmannian sources and 
$R_k$ is an infrared (IR) regulator which suppresses
 modes with momentum smaller than scale $k$. 
We require this function to satisfy conditions $R_k \to 0 \ (+\infty)$ 
as $k\to 0 \ (+\infty)$. 
Then the scale-dependent one-particle-irreducible (1PI)
 effective action $\Gamma_k[\overline{\psi},\psi]$ is defined via Legendre transformation:
\begin{equation}
\Gamma_k[\overline{\psi},\psi]
=\int^{(T)}_p\Big(\eta(p) \overline{\psi}(p)+\overline{\eta}(p)\psi(p)\Big)-W_k[\eta,\overline{\eta}]
- \int^{(T)}_p\overline{\psi}(p)R_k(p) \psi(p), \label{form04}
\end{equation}
where $\eta,\overline{\eta}$ in (\ref{form04}) are determined by inverting the relations $\delta_L W_k/\delta \eta=\overline{\psi}$ and $\delta_L W_k/\delta \overline{\eta}={\psi}$. Here, we distinguish left and right derivatives via subscripts, $\delta_L$ and $\delta_R$, respectively. Due to the property of $R_k$, $\Gamma_{k\rightarrow \infty}$ reduces to
 the classical action $S$ (up to a constant), while $\Gamma_{k=0}$ is the full 1PI quantum effective action $\Gamma$. 
$\Gamma_k[\overline{\psi},\psi]$ obeys the flow equation \cite{wetterich1993exact,Morris1,ellwanger1994flow} 
\begin{equation}
\partial_k \Gamma_k[\overline{\psi},\psi]=-{1\over 2}\mathrm{Tr}\left[{1\over \Gamma_k^{(2)}[\overline{\psi},\psi]+R_k}\p_k R_k\right], \label{intro01}
\end{equation}
where the negative sign on the right-hand side can be interpreted as the result of a closed fermionic loop, and 
$\mathrm{Tr}$ denotes the trace operation in both matrix and functional spaces with
\begin{equation}
\left(\Gamma_k^{(2)}[\overline{\psi},\psi]+R_k\right)_{p,q}
=\left({
\begin{array}{cc}
\displaystyle{\delta_L\delta_R \Gamma_k[\overline{\psi},\psi]\over \delta \overline{\psi}(p)\delta \overline{\psi}(q)}&\displaystyle{\delta_L\delta_R \Gamma_k[\overline{\psi},\psi]\over \delta \overline{\psi}(p)\delta {\psi}(q)}\\
\displaystyle{\delta_L\delta_R \Gamma_k[\overline{\psi},\psi]\over \delta {\psi}(p)\delta \overline{\psi}(q)}&\displaystyle{\delta_L\delta_R \Gamma_k[\overline{\psi},\psi]\over \delta {\psi}(p)\delta {\psi}(q)}
\end{array}}\right)+
\left(\begin{array}{cc}
0&R_k(p)\delta_{p,q}\\
-R_k(p)\delta_{p,q}&0
\end{array}\right).
\end{equation}
 Eq. (\ref{intro01}) implies that the change of the 1PI effective action in terms of $k$ is given by one-loop Feynman diagrams with dressed Green functions, 1PI interaction vertices, and a single insertion of a two-point vertex $\p_k R_k$.
  Integrating (\ref{intro01})  starting from a large enough but otherwise arbitrary scale (denoted by $\Lambda_{\rm UV}$) down to $k=0$, we can obtain $\Gamma=\Gamma_{k=0}$. 
Since we can obtain the dressed fermion propagator and 1PI vertex functions, all information on thermodynamics is available using fermionic FRG formalism. 

 To solve  the flow equation (\ref{intro01}), we consider the following vertex expansion of $\Gamma_{k}$:
\bea
\Gamma_k[\overline{\psi},\psi]
&=&\int_p^{(T)}\overline{\psi}(p) [G^{-1}-\Sigma_k](p) \psi(p) \nonumber\\
&+&\int_{p,q,q'}^{(T)}\Gamma_k^{(4)}(p; q,q')\overline{\psi}_{\up}({p\over2}+q)\overline{\psi}_{\down}({p\over2}-q)\psi_{\down}({p\over2}-q')\psi_{\up}({p\over2}+q')
+ O((\overline{\psi}\psi)^{3}) ,
\label{form06}
\eea
where  $\Sigma_k$ and  $\Gamma_k^{(4)}$ are the self-energy and the four-point vertex, respectively. 
In the following,  we truncate the expansion up to fourth order and study the flow equations of
 $\Sigma_k$ and $\Gamma_k^{(4)}$ without the introduction of auxiliary bosonic fields.
 (We call such an approach a fermionic FRG method or f-FRG method.) 
 The critical temperature $T_c$ of the superfluid transition will always be obtained from the 
 high temperature side, which means that we will work in the normal phase.  
 To study the superfluid  phase below $T_c$, one needs to  
  introduce an explicit symmetry breaking term in the classical action.
Otherwise, a second order phase transition would occur in terms of $k$ at a nonzero value (since the flow always crosses the critical surface) leading to the divergence of the four-point function and the breakdown of the flow equation  \cite{Salmhofer2004renormalization,RevModPhys.84.299}.

Let us now discuss the actual choice of the IR regulator  $R_k$, 
which needs to suppress low-energy excitations of the 
 system properly.
  In the weak coupling regime, where fermions form a Fermi sphere, 
  $R_k$ must suppress both particle and hole excitations around the Fermi
   level defined through $\bm{p}^2/2m=\mu+\sigma_0$: Here $\sigma_0$ is  a constant part of 
    the self-energy $\Sigma_0$ near the Fermi sphere. 
   A suitable regulator for this purpose is Litim's optimized regulator \cite{litim2000optimisation}
   with a self-energy term \cite{PhysRevA.81.063619};  
\begin{equation}
\int_{p}^{(T)}\overline{\psi}(p)R_k(\bm{p})\psi(p)
\equiv\int_{p}^{(T)} \overline{\psi}(p) 
\Big[{k^2\over 2m}\mathrm{sgn}(\xi(\bm{p}))-\xi(\bm{p})\Big]\theta\Big({k^2\over 2m}-|\xi(\bm{p})|\Big)\psi(p),
\label{intro03}
\end{equation}
where $\xi(\bm{p}) \equiv \frac{\bm{p}^2}{2m}-\mu-\sigma_0$ denotes
the energy relative to the Fermi level. 
For a second-order phase transition, the critical point $T=T_c$ can be determined from the Thouless criterion \cite{thouless1960perturbation} by looking at the divergence  of the   fermion-fermion scattering matrix at the total momentum 
 $p=0$. In our case, this reads 
\begin{equation}
\left[ \Gamma_{k=0}^{(4)}(p=0) \right]^{-1} =0 \ \ {\rm at}\ T=T_c.  
\label{thouless_cr}
\end{equation}

Our primary goal is to calculate the ratios $T_c/\e_F$ and $\mu/\e_F$ 
as a function of the dimensionless constant $1/(k_{\rm F} a_{\rm s})$:
Here  $a_{\rm s}$ is the scattering length between fermions, $k_{\rm F}\equiv(3\pi^2 n)^{1/3}$ with
  $n$ being the fermion number density and  $\e_F\equiv k_F^2/2m=(3\pi^2 n)^{2/3}/2m$.
Note that  the number density $n$ is related to $T$ and $\mu$ through the number equation:
\begin{equation}
n=\langle \overline{\psi}\psi \rangle = 2\int^{(T)}_p {-1\over G^{-1}(p)-\Sigma_{0}(p)}. 
\label{number_eq}
\end{equation}

\section{BCS+GMB theory from fermionic FRG method}\label{sec:Flow_Eq}

Taking the vertex expansion (\ref{form06}) of the scale dependent 1PI effective action 
$\Gamma_k[\overline{\psi},\psi]$ up to $n=2$ and 
applying the flow equation   (\ref{intro01}), 
 a closed set of equations for the  self-energy and four-point vertex  is obtained:
\begin{subequations}\label{Eq:coupled}
\be
\label{flow_se}
\p_k\Sigma_k(p)=\widetilde{\p}_k \int_l^{(T)} e^{-i l^0 0^+}
{\Gamma_k^{(4)}\left(p+l;\frac{p-l}{2},\frac{p-l}{2}\right)\over [G^{-1}-\Sigma_k+R_k](l)} ,
\ee
\bea
&&-\p_k \Gamma_k^{(4)}(p;q,q')
=\widetilde{\p}_k \Bigg[\int_l^{(T)}{\Gamma_k^{(4)}(p;q,l)\Gamma_k^{(4)}(p;l,q') \over 
[G^{-1}-\Sigma_k+R_k]\left({p\over 2}+l\right)
[G^{-1}-\Sigma_k+R_k]\left({p\over 2}-l\right)}
\label{flow_4pt}\\
&&+\sum_{\pm}\int_l^{(T)}
\scalebox{0.9}{$\displaystyle
{\Gamma_k^{(4)}\left({p\over 2}+q+l;{p/2+q-l\over 2},{p/2\mp 2q'-q-l\over 2}\right)
\Gamma_k^{(4)}\left({p\over 2}\pm q'+l;{-p/2+2q\pm q'+l\over 2},{-p/2\mp q'+l \over 2}\right)\over 
2[G^{-1}-\Sigma_k+R_k](l)[G^{-1}-\Sigma_k+R_k](q\pm q'+l)}
$}
\Bigg], 
\nonumber
\eea
\end{subequations}
where $p$ is the center-of-mass momentum, and $q,q'$ are relative momenta.
 On the right hand side of  (\ref{flow_se}) and (\ref{flow_4pt}),
 the  differential operator $\widetilde{\partial}_k$ is defined to act only on the 
 regulator $R_k$.

\begin{figure}[t]
\centering
$\partial_k \parbox{4em}{\includegraphics[width=4em]{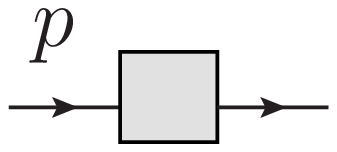}}
=\widetilde{\partial}_k \parbox{4em}{\includegraphics[width=4.5em]{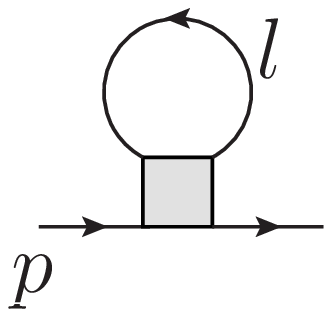}}$ \qquad \quad
$\displaystyle \partial_k \parbox{5.8em}{\includegraphics[width=5.8em]{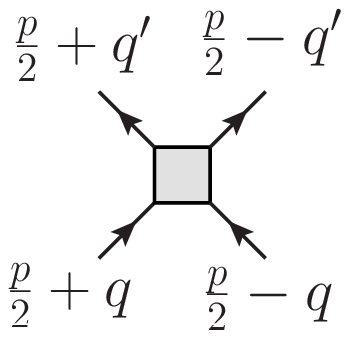}}
=\widetilde{\partial}_k\Biggl(
\parbox{5.em}{\includegraphics[width=5.em]{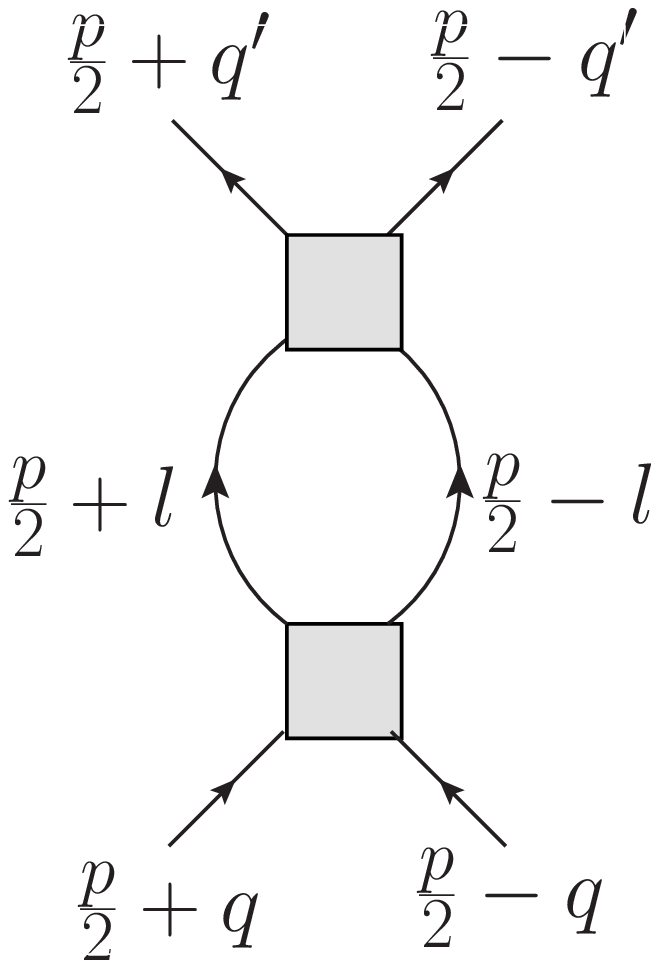}}+\sum_\pm
\parbox{7.8em}{\includegraphics[width=7.8em]{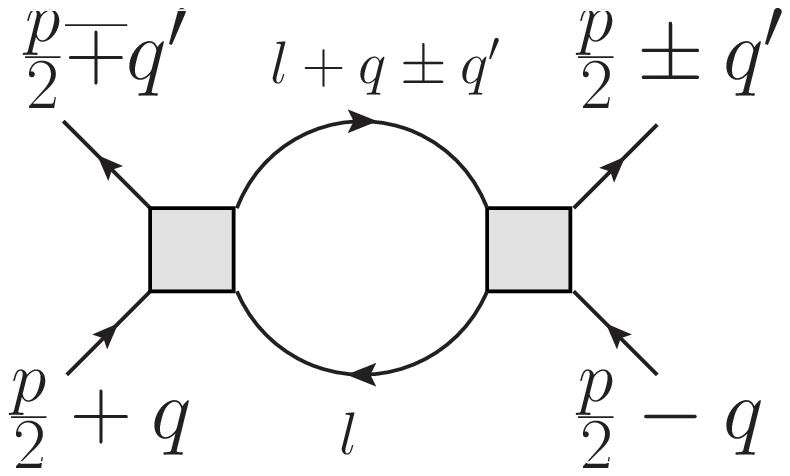}}
\Biggr)$
\caption{(a) Flow equation of the self energy $\Sigma_k$ and (b)
the four-point vertex $\Gamma_k^{(4)}$. The propagators are dressed ones at scale $k$.
\label{fig:flow_4pt}}
\end{figure}

 Diagrammatic representation of  (\ref{Eq:coupled}) is given by  Fig. \ref{fig:flow_4pt}. The first term on the ride-hand side of (\ref{flow_4pt}) describes particle-particle (PP)
  correlations, and it corresponds to the first diagram of Fig. \ref{fig:flow_4pt}(b), while the second term describes  particle-hole (PH) correlations, 
and it corresponds to the second diagram of  Fig. \ref{fig:flow_4pt}(b). Note that the PP diagram is the essential contribution forming Cooper pairs,
PH contributions will only correct the value of the transition temperature.
 In the f-FRG approach, Eqs. (\ref{Eq:coupled}) start to entangle with each 
 other throughout the FRG flow $k=\Lambda_{\rm UV} \rightarrow k=0$.

\subsection{BCS theory from f-FRG method \label{sec:rpa}}

\begin{figure}[t]
$\partial_k$\parbox{5em}{\includegraphics[width=5em]{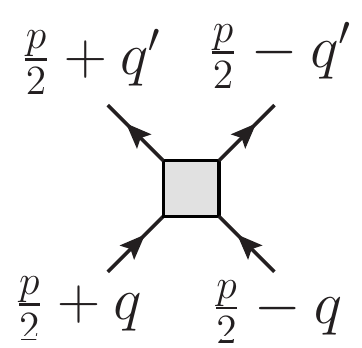}}
$\quad=\quad$ $\!\!\widetilde{\partial}_k$
\parbox{25.0em}{\includegraphics[width=5em]{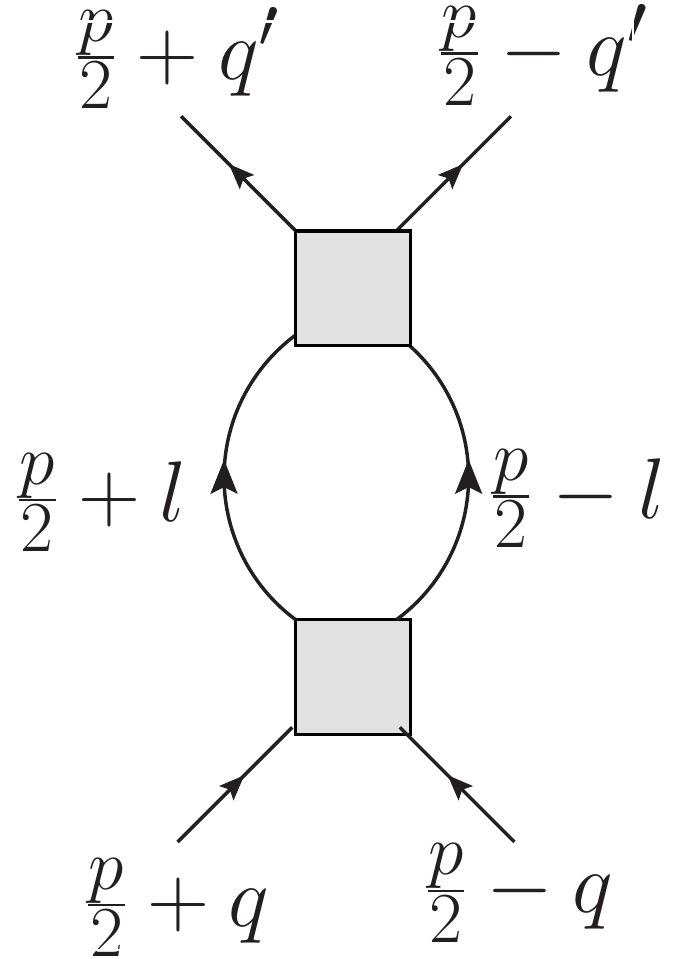}$\qquad \qquad \qquad $\includegraphics[width=0.3\textwidth]{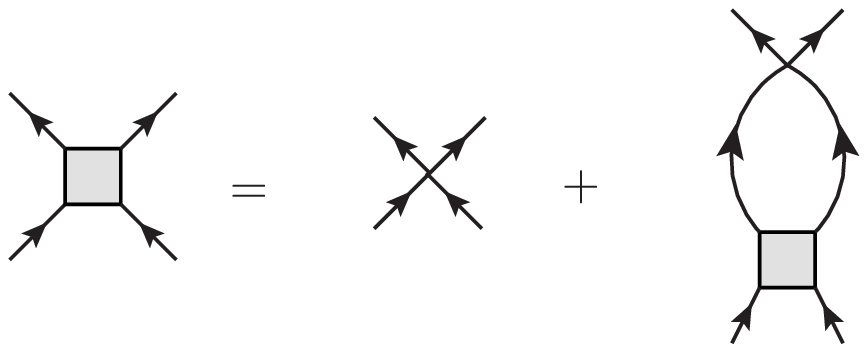}}
\caption{(a) Flow equation for $\Gamma_k^{(4)}$ (square vertex)
 restricted to the PP-channel without
 self-energy corrections, and (b) standard PP-RPA with the bare coupling $g$ denoted by 
 the point-vertex. 
 \label{fig:flow_rpa}}
\end{figure}

{ Let us first review briefly how the BCS theory is reproduced from the FRG method in the weak coupling limit 
\cite{shankar1994renormalization}. }
Let us  consider the flow of the four-point vertex ignoring the self-energy correction $\Sigma_k$. Also, 
 let us take into account only the PP correlation on the right-hand side of (\ref{flow_4pt}). Then we find 
\begin{equation}
-\p_k \Gamma_k^{(4,\mathrm{PP})}(p;q,q')=
\widetilde{\p}_k\int_l^{(T)}{\Gamma_k^{(4,\mathrm{PP})}(p;q,l)\Gamma_k^{(4,\mathrm{PP})}(p;l,q') \over 
[G^{-1}+R_k]\left({p\over 2}+l\right)[G^{-1}+R_k]\left({p\over 2}-l\right)},
\label{rpa01}
\end{equation}
where we use the label ``{\rm PP}" to identify the approximation in question.
 A diagrammatic representation of this equation is shown in Fig. \ref{fig:flow_rpa}(a). 
 Since the bare coupling in our model is a constant (see (\ref{intro02})),  $\Gamma_{k=\Lambda_{UV}}^{(4)}$ does not depend on any momenta. 
Since the flow equation (\ref{rpa01}) does not have any explicit relative momentum dependence, it cannot be produced for $\Gamma_k^{(4,\mathrm{PP})}$ either. 
 Therefore, at any $k$ it is only a function of the center of mass momentum $p$, and
  we can replace $\Gamma_k^{(4,\mathrm{PP})}(p;q,l)$ by $\Gamma_k^{\rm (4,\mathrm{PP})}(p)$.
 Eq. (\ref{rpa01}) can be rewritten as
\begin{equation}
\p_k \Bigg({1\over \Gamma_k^{\rm (4,\mathrm{PP})}(p)}\Bigg)
={\p}_k\int_l^{(T)}{1\over 
[G^{-1}+R_k]({p\over 2}+l)[G^{-1}+R_k]({p\over 2}-l)}.
\label{rpa02.5}
\end{equation}
Since in this approximation the entire $k$-dependence is due to the regulator, we have
 replaced $\widetilde{\p}_k$ on the right hand side by $\partial_k$. This means that
the flow equation can be integrated trivially as
\bea
{1\over \Gamma_k^{\rm (4,\mathrm{PP})} (p)}={1\over 
\Gamma_{\Lambda_{\mathrm{UV}}}^{\rm (4,\mathrm{PP})}}
+\int_{|\bm{l}|<\Lambda_{\rm UV}}^{(T)} {1\over [G^{-1}+R_k]({p\over 2}+l)[G^{-1}+R_k]({p\over 2}-l)} - \frac{m\Lambda_{\rm UV}}{6\pi^2}.
\label{rpa03}
\eea
Taking $k=0$ and defining 
$\frac{1}{g} \equiv \frac{1}{\Gamma^{\rm (4,\mathrm{PP})}_{\Lambda_{\rm UV}}}
-m \frac{\Lambda_{\rm UV}}{6 \pi^2} $, 
we end up with the standard BCS result above $T_c$ which
corresponds to Fig. \ref{fig:flow_rpa}(b). 
Note that, the UV cutoff $\Lambda_{\rm UV}$ in (\ref{rpa03}) can be removed 
 by the standard renormalization procedure 
 $\Gamma^{\rm (4,\mathrm{PP})}_{k=0}(0)|_{\mu=0,T=0}=\frac{4\pi a_{\rm s}}{m}$, which
  requires 
\be
{(\Gamma^{(4,\mathrm{PP})}_{\Lambda_{\mathrm{UV}}}})^{-1} 
= {m\over 4\pi a_{\rm s}}-\left(\int_0^{\Lambda_{\mathrm{UV}}} {\diff^3\bm{l}\over (2\pi ^3)}{1\over 2\bm{l}^2/2m}-{m\Lambda_{\mathrm{UV}}\over 6\pi ^2}\right)
= {m\over 4\pi a_{\rm s}}-{m\Lambda_{\mathrm{UV}}\over 3\pi^2}.
\ee

 After renormalization, the Thouless criterion 
  $[\Gamma_{k=0}^{(4)}(p=0)]^{-1} \big|_{T=T_c}=0$ together with (\ref{rpa03}) 
  leads to the standard form
\begin{equation}
{\pi\over 2|a_{\rm s}|}={1\over 2}\int_0^{\infty}\sqrt{2m\e}\diff\e
\left[{\tanh\left({(\e-\mu)/ 2 T_c}\right)\over \e-\mu}-{1\over \e}\right].
\label{rpa04}
\end{equation}
If we further take 
 the weak coupling limit ($(k_F a_{\rm s})^{-1}\to - \infty$) with
 $\mu=\e_F$, (\ref{rpa04}) provides the standard BCS result
\bea 
\frac{T^{\mathrm{BCS}}_c}{\e_F}
\equiv \frac{8}{\pi} e^{\gamma_{\rm E}-2}
e^{ -\pi/ 2k_F |a_{\rm s}| }.
\label{BCSres}
\eea

\subsection{GMB correction from f-FRG method\label{sec:gmb}}

Now we consider the particle-hole correlation (second term on the right hand side of Fig. \ref{fig:flow_4pt}) 
in the flow equation of $\Gamma_k^{(4)}$. As reported in Ref. \cite{PhysRevB.78.174528}, this leads to the Gorkov and Melik-Barkhudarov (GMB) correction  
\cite{gorkov1961contribution}.
The main aim of this subsection is to clarify the origin of the GMB correction in detail
 and to analyze the approximations to be employed in a precise way. Such an analysis is particularly useful for
  going beyond the BCS+GMB theory in later sections.

In general, the four-point vertex function $\Gamma_k^{(4)}(p;q,q')$ depends not only 
on the center-of-mass momentum $p$ but also on relative momenta $q$ and $q'$. 
If we restrict ourselves to particle-particle correlations, the dependence on $q$ and $q'$
 disappears as we have shown in the previous
 subsection. However, once we have particle-hole correlations, such a simplification does not
 occur anymore. However, since we are interested in s-wave superfluidity, instead of solving the full momentum dependence, we can consider
the case $|\bm{q}|=|\bm{q}'|=k_F$ and take the s-wave projection averaged over the directions 
  of $\bm{q}$ and $\bm{q}'$: 
$\Gamma_k^{\rm (4)}(p;q^0,q'^{0})=\int{\diff^2\hat{\bm{q}}\over 4\pi}{\diff^2\hat{\bm{q}}'\over 4\pi} \left. \Gamma_k^{\rm (4)}(p;q,q')\right|_{|\bm{q}|=|\bm{q}'|=k_F}$. This is motivated by
 the fact that, for low energy single-particle excitations are given by fermionic quasi-particles 
in the vicinity of the Fermi surface, and $\bm{q}$ and $\bm{q}'$ 
 will be of the order of the Fermi momentum $k_F\approx \sqrt{2m\mu}$
  with the loop momentum $\bm{l}$ being also restricted to the region $|\bm{l}|$ $\sim k_F$
  due to the presence of the regulator $R_k$.
  Finally, we define $\Gamma_k^{(4)}(p)\equiv \Gamma_k^{(4)}(p;0,0)$, 
 so that $\Gamma_k^{(4)}(p=0)$ can be regarded as the effective coupling constant at scale $k$.
 The corresponding flow equation  is given by 
\begin{eqnarray}
&&\p_k \Bigg({1\over \Gamma_k^{(4,\mathrm{PP+PH})}(0)}\Bigg)=
{\p}_k\int_{l}^{(T)}{1\over [G^{-1}+R_k](l)[G^{-1}+R_k](-l)}
\nonumber\\
&&+{\p}_k
\int\limits_{\{|\bm{q}|=k_F\}}
{\diff^2\hat{\bm{q}}\over 4\pi}{\diff^2\hat{\bm{q}}'\over 4\pi}
\int_{l}^{(T)}\hspace{0.2em}{1
\over [G^{-1}+R_k](l) [G^{-1}+R_k](q-q'+l)},
\label{gmb03}
\end{eqnarray}
where we have used the label "PP+PH" to identify the approximation taken here. 
Note that, the angular integration of $\hat{\bm{q}}$ and $\hat{\bm{q}}'$ 
 still remain in the particle-hole contribution through the regularized fermion propagator.
{ Further discussions on the  approximation
 adopted here  are  given at the end of this subsection
  on the basis of the detailed flow pattern as a function of $k$ in the weak coupling regime.} 
  
Since both sides of (\ref{gmb03}) became total derivatives, it can be solved as
\bea
&&{1\over\Gamma_k^{(4,\mathrm{PP+PH})}(0)}={1\over \Gamma_k^{(4,\mathrm{PP})}(0)}\nonumber\\
&&+\int\limits_{0}^{8m\mu}{\diff |\bm{Q}|^2\over 8m\mu}\int{\diff^3\bm{l}\over (2\pi)^3}{n_F\left({(\bm{l}+\bm{Q})^2\over 2m}-\mu+R_k(\bm{l}+\bm{Q})\right)-n_F\left({\bm{l}^2\over 2m}-\mu+R_k(\bm{l})\right) \over \left({(\bm{l}+\bm{Q})^2\over 2m}-\mu+R_k(\bm{l}+\bm{Q})\right)-\left({\bm{l}^2\over 2m}-\mu+R_k(\bm{l})\right)},\label{gmb04}
\eea
where  $\bm{Q}\equiv\bm{q}-\bm{q'}$ and $n_F$ is the Fermi-Dirac distribution function. 
The first term in the right-hand side of (\ref{gmb04}) has already been evaluated in  (\ref{rpa03}).
Also, considering the fact that  the critical temperature $T_c$ 
is much smaller than the Fermi energy $\e_F$ in the weak coupling regime, 
the second term in the right-hand side of (\ref{gmb04}) can be evaluated at $T=0$ as  
\be
\int\limits_{0}^{8m\mu}{\diff |\bm{Q}|^2\over 8m\mu}\int{\diff^3\bm{l}\over (2\pi)^3}{\theta\left({(\bm{l}+\bm{Q})^2\over 2m}-\mu\right)-\theta\left({\bm{l}^2\over 2m}-\mu\right) \over \left({(\bm{l}+\bm{Q})^2\over 2m}-\mu\right)-\left({\bm{l}^2\over 2m}-\mu\right)}=-{1+\ln 4\over 3}{m k_F\over 2\pi^2}. 
\ee
Then we find that, at the end of the flow ($k\rightarrow 0$),
 the particle-hole contribution simply shifts 
 the value of the inverse scattering length from $a_s$ to $a_s^{\rm eff}$:
\be 
 \frac{1}{a_s^{\rm eff}} =  \frac{1}{a_s}-\frac{2k_F}{\pi}\frac{1+\ln 4}{3}.
\ee 
Therefore, taking into account the correction coming from PH correlations, the BCS critical temperature in the weak coupling regime is corrected as
\begin{equation}
\frac{T_c^{\mathrm{GMB}}}{\e_F}=\frac{1}{(4e)^{1/3}}\cdot \frac{T_c^{\mathrm{BCS}}}{\e_F}.
\label{gmb05}
\end{equation}
This is exactly the Gorkov and Melik-Barkhudarov result\cite{gorkov1961contribution,PhysRevLett.85.2418}, showing the reduction of $T_c$ due to the effect of order parameter fluctuations in agreement with previous FRG studies as well \cite{PhysRevB.78.174528}. 

\begin{figure}
\includegraphics[scale=0.55]{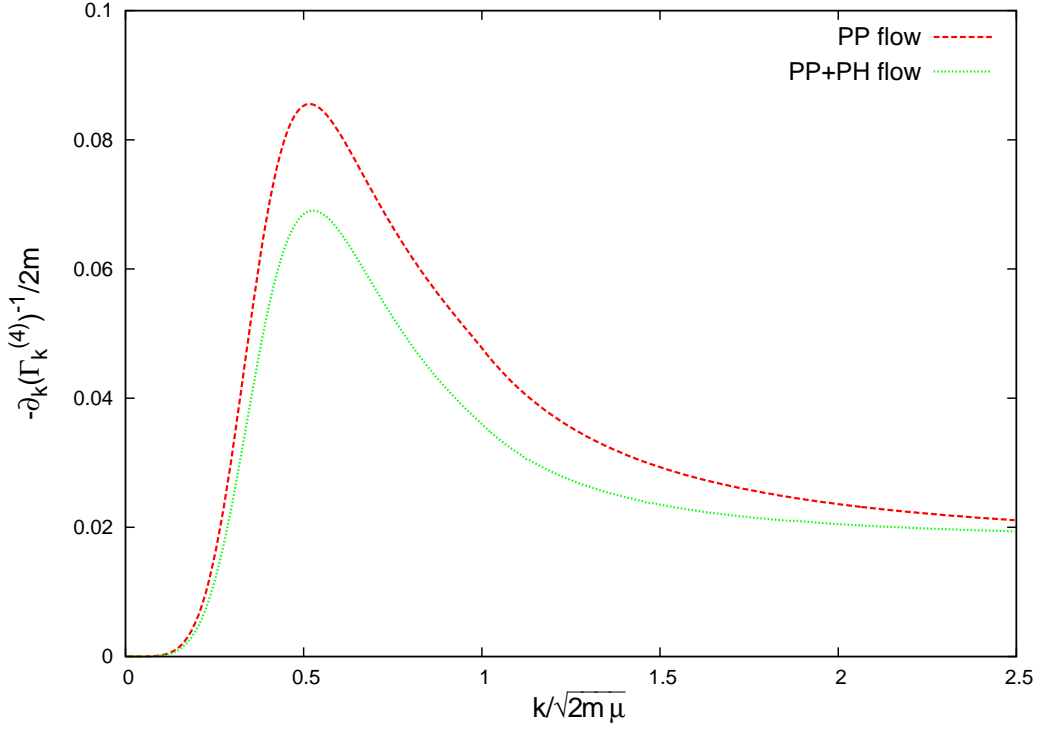}
\caption{Derivative of $1/\Gamma_k^{(4)}(0)$ at $T_c/\mu=0.06$ (which corresponds to $(|k_Fa_s|)^{-1} \sim {\mathcal O}(1)$) with and without the PH loop included. The figure shows that only when $k \sim k_F \approx \sqrt{2m\mu}$, particle-hole fluctuations change the behavior of the flow. }
\label{Fig3}
\end{figure}

\begin{table}
\centering
\begin{tabular}{c|c|c|c}\hline
&$\quad k^2/2m\ll  \pi T_c \quad$ & $\quad \pi T_c\ll k^2/2m\ll k_F^2/2m \quad$ & $\quad k^2/2m\gg k_F^2/2m\quad$ \\ \hline
PP & $\mathcal{O}(k_F k^5/T_c^3)$ & $\mathcal{O}(k_F/k)$ & $\mathcal{O}(1)$ \\
PH & $\mathcal{O}(k^5/k_F T_c^2)$ & $\mathcal{O}(k/k_F)$ & $\mathcal{O}(k_F^3/k^3)$ \\ \hline
\end{tabular}
\caption{\label{tab0} Contributions from the particle-particle (PP) correlation  and the
 particle-hole (PH) correlation to $\p_k (\Gamma_k^{(4)}(0))^{-1}$ in the right-hand side of  Eq. (\ref{gmb03}). This behavior is consistent with that of the prediction from Shanker's RG analysis \cite{shankar1994renormalization}. }
\end{table} 

Let us now discuss the structure of the flow equation (\ref{gmb03}) in more detail
 to gain a deeper understanding of why exactly the GMB result was reproduced from f-FRG within the above approximation. 
The parametric $k$-dependence of the first term (particle-particle correlation)  and the 
second term (particle-hole correlation) of the right-hand side of (\ref{gmb03})  are summarized in Table \ref{tab0}
(the corresponding explicit formulas are given in \ref{app:asymp}). 
 We mention here that PH contribution has opposite sign to the PP contribution.
 From Table \ref{tab0}, we observe that  PH contribution is much smaller than the PP contribution in the high-energy region ($k\gg k_F$) and the low-energy region ($k\ll k_F$). 
\footnote{The behavior of these terms in
 f-FRG method  are qualitatively consistent with the Shankar's RG analysis of interacting fermions \cite{shankar1994renormalization}. }
Only for $k \simeq k_F$ do the PH and PP contributions become comparable.  Such a behavior can be seen explicitly in Fig. \ref{Fig3}, where  numerical solutions of $\p_k (\Gamma_k^{(4)}(0))^{-1}$ with and without the PH correlation are plotted. The figure shows that the screening of the effective coupling from the PH correlation takes place  only around $k\sim k_F$. 
     This implies that we can neglect the momentum dependence of $\Gamma_k^{(4)}$ in the PH contribution, at least for weak couplings, where we have the hierarchy $1/a_s\gg k_F\gg T_c/k_F$. In such a regime, $1/\Gamma_k^{(4)}$ is $\mathcal{O}(1/a_s)$ around $k\sim k_F$, therefore the momentum dependence of $\Gamma_k^{(4)}$ in the PH contribution is negligible. 
   Although such a justification is questionable beyond the weak coupling limit,
  we will later adopt this approximation as a working hypothesis
  to make a qualitative study towards the unitary regime.
  
For later purposes, we show here the structure of $\Gamma_k^{(4,\rm{PP+PH})}$ as a function of the center-of-mass momentum $p$ for large values of the flow parameter $k$.
  In such an asymptotic regime,  PP correlation is the dominant contribution
  as discussed above. Therefore, (\ref{rpa03}) leads to 
\begin{equation}
{1\over \Gamma_k^{(4,\mathrm{PP+PH})}(p)} \xrightarrow{k \to\infty} -{2m\over 6\pi^2}\left[k-{3\pi \over 4 a_s}+{2m\over k}\left(i p^0+{\bm{p}^2\over 4m}-{3\over 2}\mu\right)\right]. 
\label{rpa04a}
\end{equation}

\section{f-FRG  with fermion self-energy}\label{sec:self_energy}

So far, we have not taken into account the effect of the self-energy correction $\Sigma_k$ in 
the flow equations.  Since $\sigma_k$ (the constant part of $\Sigma_k$ near the Fermi sphere) 
gives a shift of the chemical potential $\mu \rightarrow \mu_k\equiv \mu + \sigma_k$,
 its importance grows as the system approaches the unitary regime.
 In the following, we focus only on the lowest Matsubara frequency part of $\Sigma_k$ 
and define $\sigma_k$ as its real part:
$\sigma_k \equiv \Re \Sigma_k(\pm \pi T,\bm{0})$.
We note that even with the present (constant) self-energy, non-trivial resummation of the 
original perturbative series occurs via the self-consistent nature of the coupled flow equations, and therefore 
it is a good starting point for analysing the role of the self-energy going beyond BCS+GMB theory.
 In this section we first will study the behavior of $\sigma_k$ for asymptotically large $k$, and formulate coupled flow equations of  
 $\sigma_k$ and $\Gamma_{k}^{(4)}$, an then proceed to perform a numerical solution of them.

\subsection{Asymptotic behavior of the self-energy at large $k$}

Within the momentum independent vertex discussed in the previous section, using (\ref{flow_se}), the constant part of the self-energy near the Fermi sphere (i.e. $\sigma_k$),
satisfies
\bea
\label{Eq:sigma_k2}
\partial_k \sigma_k= \tilde{\partial}_k \int_l^{(T)} \frac{\Gamma_k^{(4)}(0)}{G^{-1}(l)
-\sigma_k+R_k(l)}, 
\eea
which can be rewritten using an effective Fermi level $\mu_k=\mu+\sigma_k$ as 
\bea
\partial_k \mu_k&=&-\frac{(2m)^{1/2}k\Gamma_k^{(4)}(0)}{3\pi^2}\left[\Big((\mu_0+k^2/2m)^{3/2}-(\mu_0)^{3/2}\Big)n_F'(\omega_{+})\right.\nonumber\\
&&\left.-\Big((\mu_0)^{3/2}-\Re(\mu_0-k^2/2m)^{3/2}\Big)n_F'(\omega_{-})\right],
\label{Eq:coupled-a}
\eea
where $\omega_\pm = \pm k^2/2m + \mu_0-\mu_k$, and $n_F$ is the Fermi-Dirac distribution function. 
By taking $k\rightarrow \infty$ (note that $k a_s$ can take any negative values) and using the asymptotic form of $\Gamma_k^{(4)}(0)$ given in
(\ref{rpa04a}), one finds
\bea
\sigma_k = e^{-k^2/2mT}\left({k^2/2m\over 1-3\pi/4k a_s}+\mathcal{O}(1)\right).
\label{eq:wrong_asymp}
\eea
This asymptotic behavior is, however, not correct as can be seen from the following argument.
 First of all, the present theory is asymptotically free in the sense that
 $\Gamma_k^{(4)}(0) \simeq -(3\pi^2)/(m(k-3\pi/4 a_s))$ for large $k$ as obtained from (\ref{rpa04a}).
 Then, since perturbative analysis is valid for large $k$, we can evaluate $\sigma_{k}$ for large $k$ as
  \bea
\sigma_k \simeq \Re\ \int_l^{(T)} \frac{\Gamma^{(4)}_k(p+l)e^{-il^00^+}}{i(p^0+l^0)+\frac{\bm{l}^2}{2m}-\mu+R_k(\bm{l})},
\label{Eq:sigma_appr}
\eea
{ where $p^0=\pm \pi T$.} 
Taking into account the correct leading-order frequency dependence  
as  $\displaystyle \Gamma_k^{(4)}(l) \simeq -\frac{(3\pi^2 k/2m^2)}{(il^0+k^2/2m-3\pi k/8m a_s)}$ and carrying out the integral,
one finds
\bea 
\sigma_k &=& \Re \ \ \int_l^{(T)} {e^{-i l^0 0^+}\over i l^0+{\bm{l}^2\over 2m}-\mu+R_k(\bm{l}) }{-3\pi^2 k/2m^2 \over i (p^0+l^0)+k^2/2m-3\pi k/8m a_s}\nonumber\\
&\simeq& \frac{\sqrt{2m}(\mu+\sigma_0)^{3/2}}{2k(1-3\pi/8k a_s)},
\label{se_rpa02}
\eea
where we omitted the spatial-momentum dependence of $\Gamma_k(p+l)$ giving only subleading contributions.
Thus we observe that $\sigma_k$ must decrease as  $1/k$, unlike (\ref{eq:wrong_asymp}).
Another way to see the problem of (\ref{eq:wrong_asymp}) is that
 the number density $n_k=\langle \overline{\psi}\psi\rangle$  
 for large $k$ simply vanishes contrary to the 
 true behavior  expected from  the lowest-order perturbation at large $k$;
$n_{k} \rightarrow \big( 2m\mu_0\big)^{3/2}/(3\pi^2)$.

\subsection{Coupled flow equations for self-energy and 4-point vertex}  

From the discussion of the previous subsection, we see that 
the momentum dependence of $\Gamma_k^{(4)}$, especially its frequency dependence, is 
essential to determine $\sigma_k$ in the f-FRG approach. 
To study such a frequency dependence by avoiding the
quite demanding numerical calculation of solving the coupled flow-equation of $\Sigma_k$ and $\Gamma_k$ with full momentum 
dependence, we adopt a hybrid approach described below as a first step, in order to explore the 
 effect of the self-energy.

 Let us  start with the following coupled equations;
\bea
\label{eq:coup_sigma-flow}
\partial_k \sigma_k&=&\Re\tilde{\partial}_k \int_l^{(T)} \frac{\Gamma_k^{(4)}(\pm\pi T+l^0,{\bm{l}})}{
G^{-1}(l)-\sigma_k+R_k(l)}, \\
\label{eq:coup_gamma-flow}
\partial_k \Gamma_k^{(4)-1}(0)&=&\tilde{\partial}_k \int_l^{(T)} \frac{1}{[G^{-1}(l)
-\sigma_k+R_k(l)][G^{-1}(-l)-\sigma_k+R_k(-l)]}  \nonumber\\
&+&\tilde{\partial}_k \int\limits_{0}^{8m\mu}\frac{\diff |\bm{Q}|^2}{8m\mu} \int_l^{(T)} \frac{1}{[G^{-1}-\sigma_k+R_k](l)[G^{-1}-\sigma_k+R_k](Q+l)},
\eea
with  $Q=(0,\bm{Q})$.  
In (\ref{eq:coup_sigma-flow}),  the frequency of the fermion self-energy  is restricted to the lowest values
$\pm \pi T$. 
After performing the Matsubara sum, $\sigma_k$ and $\sigma_0$ appear as a combination $\sigma_0-\sigma_k$  $(=\mu_0-\mu_k)$.
Since the approximate particle-hole symmetry would make  this combination small for $k < k_F$  where PH contribution is already not significant,  we take $\sigma_k=\sigma_0$ in the
  actual calculation of the  PH contribution.
  
To take into account the momentum dependence in a minimal, but sufficient way we make the following
expansion and keep first few terms:
\bea
\label{Eq:expan}
\Gamma_k^{(4)-1}(p^0+l^0,{\bm{l}})
 \approx -Z_k^{-1} \left[ i(l^0 + p^0)+ S_k^{(1)} \cdot |{\bm{l}}| + S_k^{(2)}\cdot {\bm{l}}^2+|\mu^B_k|\right] ,\eea
where we have introduced the notations
\bea
&&Z_k^{-1}=i\partial_{l^0}\Gamma_k^{(4)-1}(0), 
\quad |\mu_k^B|=-Z_k \Gamma_k^{(4)-1}(0), \nonumber\\
&& S_k^{(1)}=-Z_k \partial_{|\bm{l}|} \Gamma_k^{(4)-1}(0), 
\quad S_k^{(2)}=-Z_k \partial_{|\bm{l}|^2}\Gamma_k^{(4)-1}(0).
\eea
Now we adopt a hybrid approach in which 
 $\Gamma_k^{-1}(0)$ is calculated from the flow equation, while its derivatives with respect
 to the frequency and momentum are estimated by the  RPA analysis in the particle-particle channel given 
  in Appendix A. 
  After applying the expansion (\ref{Eq:expan}), the flow equation for the self-energy becomes
\bea
\label{Eq:flow_sigma_a}
\partial_k \sigma_k&=&\Re\tilde{\partial}_k \int_l^{(T)} \frac{1}
{G^{-1}(l)-\sigma_k+R_k(l)}\frac{-Z_k}{i(l^0 \pm \pi T)+S_k^{(1)}|\bm{l}|+S_k^2\bm{l}^2+|\mu_k^B|} \\
\label{Eq:flow_sigma_b}
&=&-\Re\tilde{\partial}_k\int_0^{\infty}\frac{dll^2}{2\pi^2}\Big(n_B(\omega_k^B(\bm{l}))+n_F(\omega_k(\bm{l}))\Big)
\frac{Z_k}{\omega_k(\bm{l})-\omega_k^B(\bm{l})\mp i \pi T},
\eea
where $\omega_k(\bm{l})=\frac{\bm{l}^2}{2m}-\mu_k+R_k({\bm{l}})$, and $\omega_k^B(\bm{l})=|\mu_k^B|+S_k^{(1)}|{\bm l}|+S_k^{(2)}{\bm l}^2$, with $n_B$ being the Bose-Einstein distribution function. 

This formula shows the importance of the momentum dependence of $\Gamma_k^{(4)}$ as well as  
its correct boundary condition in the frequency space. Due to these,
 bosonic (Cooper pair) contributions appeared in the flow equation of the self-energy.
 We note that a  peculiar linear term in $\omega_k^B(\bm{l})$
  is only due to the presence of the regulator, and one can show that it disappears at $k=0$. 
   After performing the $\tilde{\partial}_k$ differentiation, we obtain the following coupled flow equations:
\begin{subequations}
\bea
\label{Eq:flow_sigma_c}
&&\partial_k \mu_k=-{k\over 2m}\int\limits_{\sqrt{2m\mu_0}}^{\sqrt{2m\mu_0+k^2}} \frac{dl}{\pi^2}l^2 Z_k \Re\Bigg[\frac{\partial_\omega n_F(\omega_{k+})}{\omega_{k+}-\omega_k^B(\bm{l})\mp i\pi T}-\frac{n_B(\omega_k^B(\bm{l}))+n_F(\omega_{k+})}{(\omega_{k+}-\omega_k^B(\bm{l})\mp i \pi T)^2}\Bigg] \nonumber\\
&&+{k\over 2m}\int\limits_{\Re\sqrt{2m\mu_0-k^2}}^{\sqrt{2m\mu_0}} \frac{dl}{\pi^2}l^2 Z_k \Re\Bigg[\frac{\partial_\omega n_F(\omega_{k-})}{\omega_{k-}-\omega_k^B(\bm{l})\mp i\pi T}-\frac{n_B(\omega_k^B(\bm{l}))+n_F(\omega_{k-})}{(\omega_{k-}-\omega_k^B(\bm{l})\mp i \pi T)^2}\Bigg], 
\eea
\bea
&&\partial_k \Gamma^{(4)-1}_k= \frac{(2m)^{1/2}k}{3\pi^2}\Big(-\frac{1}{2\omega_{+}^2}+\frac{n_F(\omega_{+})}{\omega_{+}^2}-\frac{n_F'(\omega_{+})}{\omega_{+}}\Big)\Big((\mu_0+k^2/2m)^{3/2}-(\mu_0)^{3/2}\Big)\nonumber\\
&&-\frac{(2m)^{1/2}k}{3\pi^2}\Big(-\frac{1}{2\omega_{-}^2}+\frac{n_F(\omega_{-})}{\omega_{-}^2}-\frac{n_F'(\omega_{-})}{\omega_{-}}\Big)\Big((\mu_0)^{3/2}-\Re(\mu_0-k^2/2m)^{3/2}\Big) \nonumber \\
&&+\widetilde{\partial}_k  \int\limits_{0}^{8m\mu}{\diff |\bm{Q}|^2\over 8m\mu}\int{\diff^3\bm{l}\over (2\pi)^3} \scalebox{0.95}{$\displaystyle \frac{n_F\Big( {({\bm{l}})^2\over 2m}+R_k({\bm{l}})-\mu_k\Big)-n_F\Big( {({\bm{l}}+{\bm{Q}})^2\over 2m}+R_k({\bm{l}}+{\bm{Q}})-\mu_k\Big)}{({\bm{l}})^2/2m+R_k({\bm{l}})-({\bm{l}}+{\bm{Q}})^2/2m-R_k({\bm{l}}+{\bm{Q}})}$},
\label{Eq:coupled-b}
\eea
\end{subequations}
where $\omega_{k\pm}=\pm k^2/2m + \mu_0 -\mu_k$. The effective Fermi level $\mu_k$ starts to flow from the chemical potential at $k= \infty$, $\mu_{k=\infty}=\mu$, and converges to  $\mu_0=\mu+\sigma_0$. 
Because of the appearance of bosonic (Cooper pair) contributions to the flow equation for $\mu_k=\mu+\sigma_k$, 
(\ref{Eq:flow_sigma_c}) indeed shows the correct asymptotic behavior at large $k$ consistent with the 
 discussion in the previous subsection:
\be
\partial_k \mu_k = -\frac{\sqrt{2m}\mu_0^{3/2}}{2(k-3\pi/8 a_s)^2} + {\mathcal O}(1/k^4) 
 \Longrightarrow \sigma_k = \frac{\sqrt{2m}(\mu+\sigma_0)^{3/2}}{2k(1-3\pi/8k a_s)} + {\mathcal O}(1/k^3). 
\ee

\section{Numerical Results}\label{sec:results}

\begin{figure}[tbp]
\includegraphics[scale=0.55]{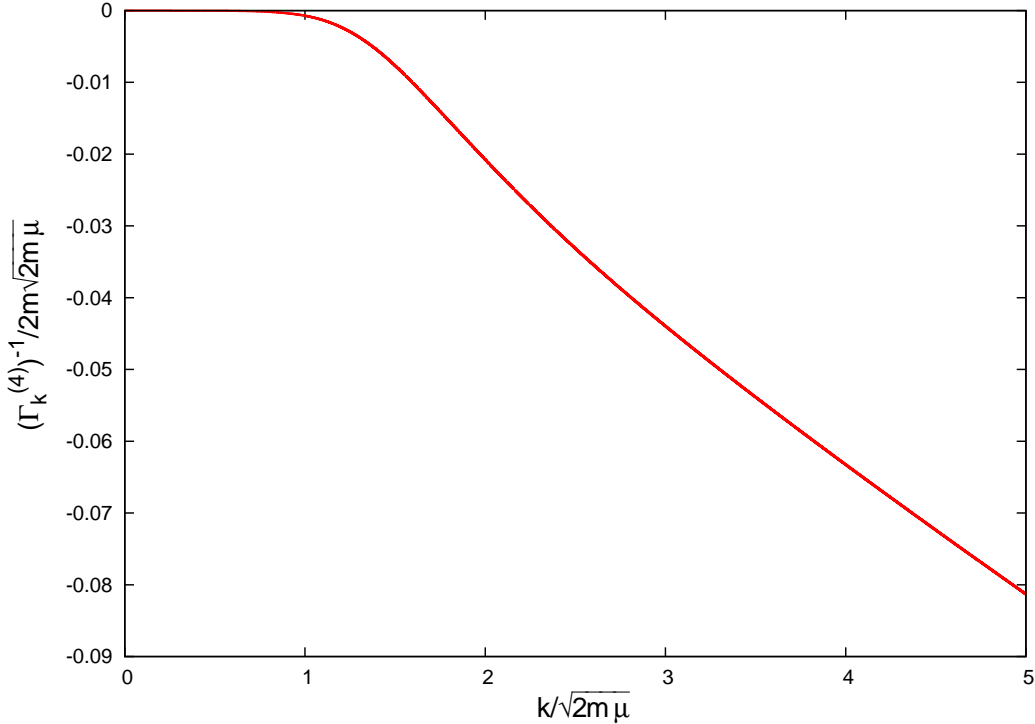}
\caption{Flow of the inverse of the four-point function at unitarity in PP+PH+SE approximation. The plot demonstrates that the singularity only appears at $k=0$.}
\label{Fig_new}
\end{figure}  

 \begin{figure}[tbp]
\includegraphics[scale=0.55]{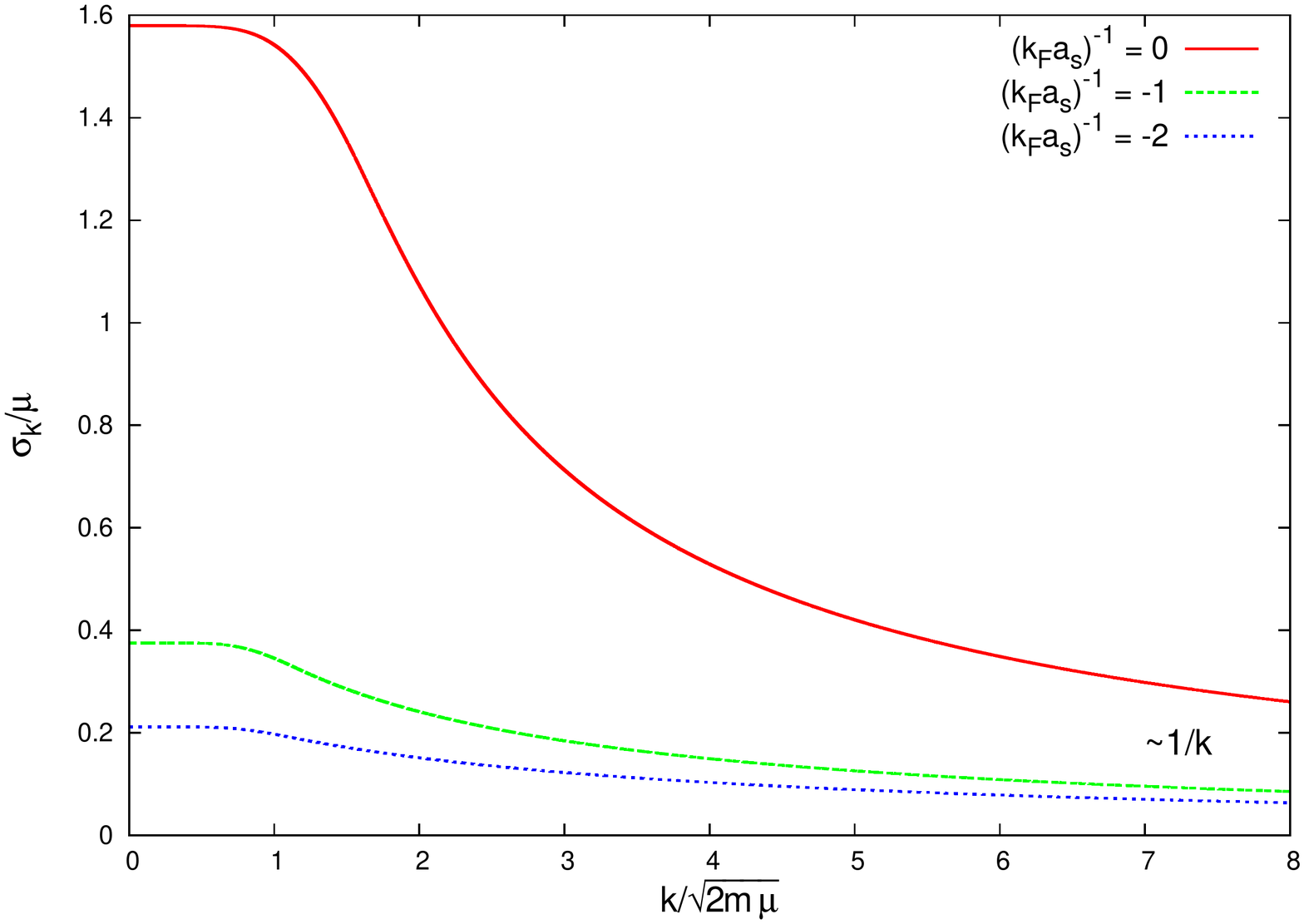}
\caption{Self-energy $\sigma_k=\mu_k-\mu$ as a function of $k$ for $(k_F a_s)^{-1}=-2,-1,0$ at $T=T_c$.}
\label{Fig4}
\end{figure}  
  Now we are ready to solve (\ref{Eq:flow_sigma_c}) and (\ref{Eq:coupled-b}) together with the  number equation, (\ref{number_eq}).
In order to get the transition temperature $T_c$, one may start from a large $k$ at the   UV cutoff scale, and let the flow equations run towards $k=0$. Then $T_c$ is obtained through the Thouless criterion, $\Gamma^{(4)-1}_{k=0}(0)=0$,
 while the chemical potential $\mu$ and the number density $n$ are related via the number equation (\ref{number_eq}). 
  However, the actual numerical calculation is much more efficient by  starting  from $k=0$ with $\Gamma_{k=0}^{(4)-1}(0)=0$ and letting the flow towards large $k$.
 Keeping the temperature on a certain value and choosing $\mu=1$, the flow of $\sigma_k$ and $\Gamma_k^{(4)-1}$ are numerically obtained on a grid with a step size of $\Delta k = 10^{-4}$. Then we fit these quantities  by their known asymptotic behavior for large $k$  to determine $\sigma_{k=\infty}$ and $\Gamma_{k=\infty}^{(4)}$. 
  We choose the highest scale to be $k_{UV}=50.0$ and fit the functions in the interval $[49.0,50.0]$. 
  After subtracting the ``divergent'' $-mk/3\pi^2$ piece in (\ref{rpa04a}), $\Gamma^{(4)-1}_{k=\infty}$ should be  equal to $m/4\pi a_s$, from which we obtain the function $a_s=a_s(T_c)$. 
  Since the self-energy always has to approach zero at large $k$, a unique solution is obtained by adjusting $\sigma_{k=0}$ this way.

Even after taking into account the self-energy correction, we found numerically that the flow of the four-point vertex $\Gamma_k^{(4)}$ shows a qualitatively similar behavior to the PP+PH flow shown in Fig. \ref{Fig3}. Also, the scalings with respect to flow parameter $k$ are the same as those given in Table \ref{tab0}.  Therefore, we can safely use the Thouless criterion (\ref{thouless_cr}) to determine the critical temperature of the superfluid transition without singular behavior of the flow at $k>0$. We show a typical flow of $[\Gamma_{k}^{(4)}(0)]^{-1}$ in
Fig. \ref{Fig_new}, which indicates that $[\Gamma_{k}^{(4)}(0)]^{-1}$ increases monotonically as $k$ decreases.

\begin{figure}
\includegraphics[scale=0.55]{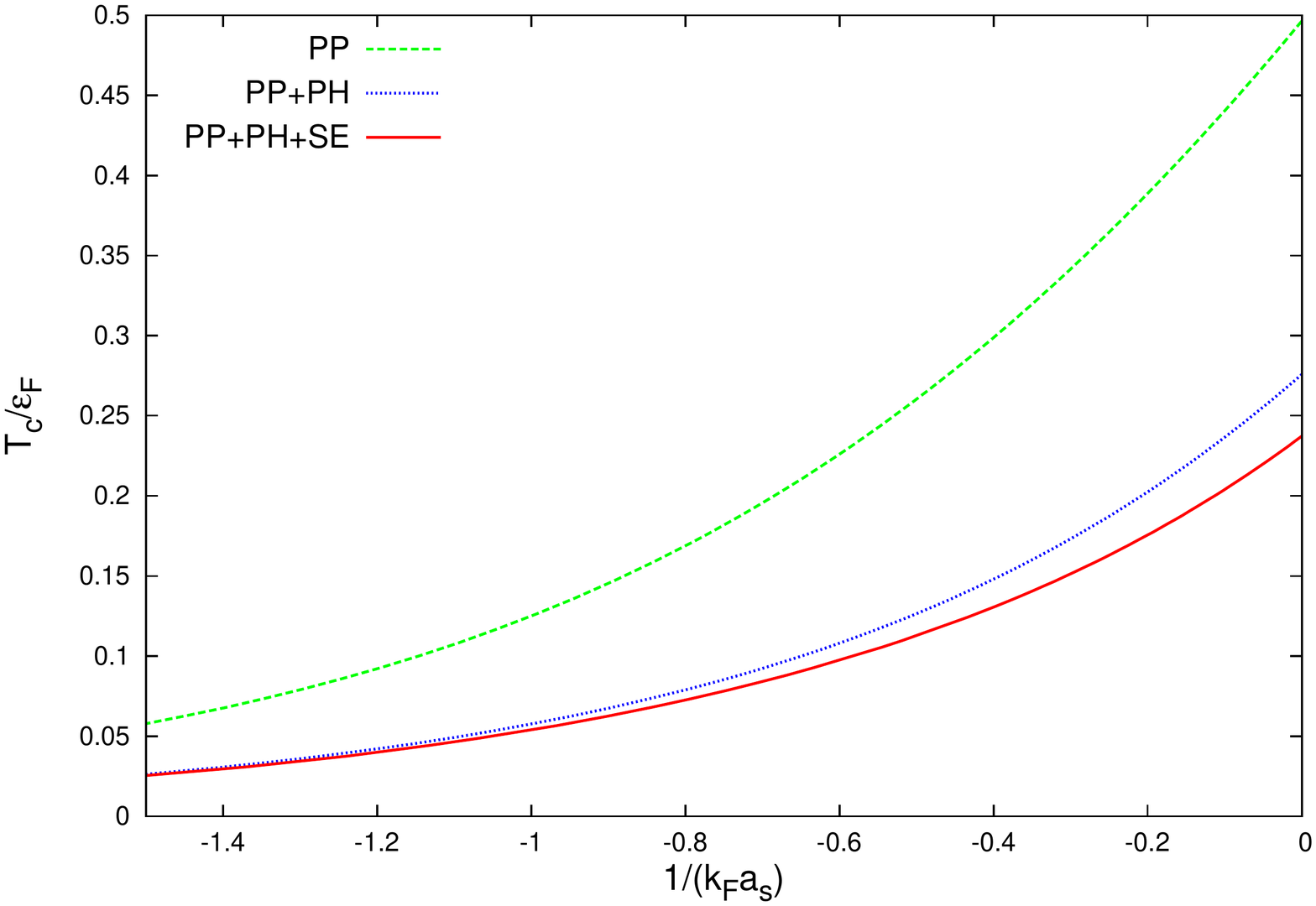}
\caption{ $T_c/\e_F$ as a function of the dimensionless scattering strength, $(k_F a_s)^{-1}$, 
in different levels of approximation: 
 PP (particle-particle correlation only), PP+PH (particle-particle and particle-hole correlation
 without self-energy), PP+PH+SE (particle-particle and particle-hole correlation with self-energy).
} 
\label{Fig5}
\end{figure}

\begin{figure}
\includegraphics[scale=0.55]{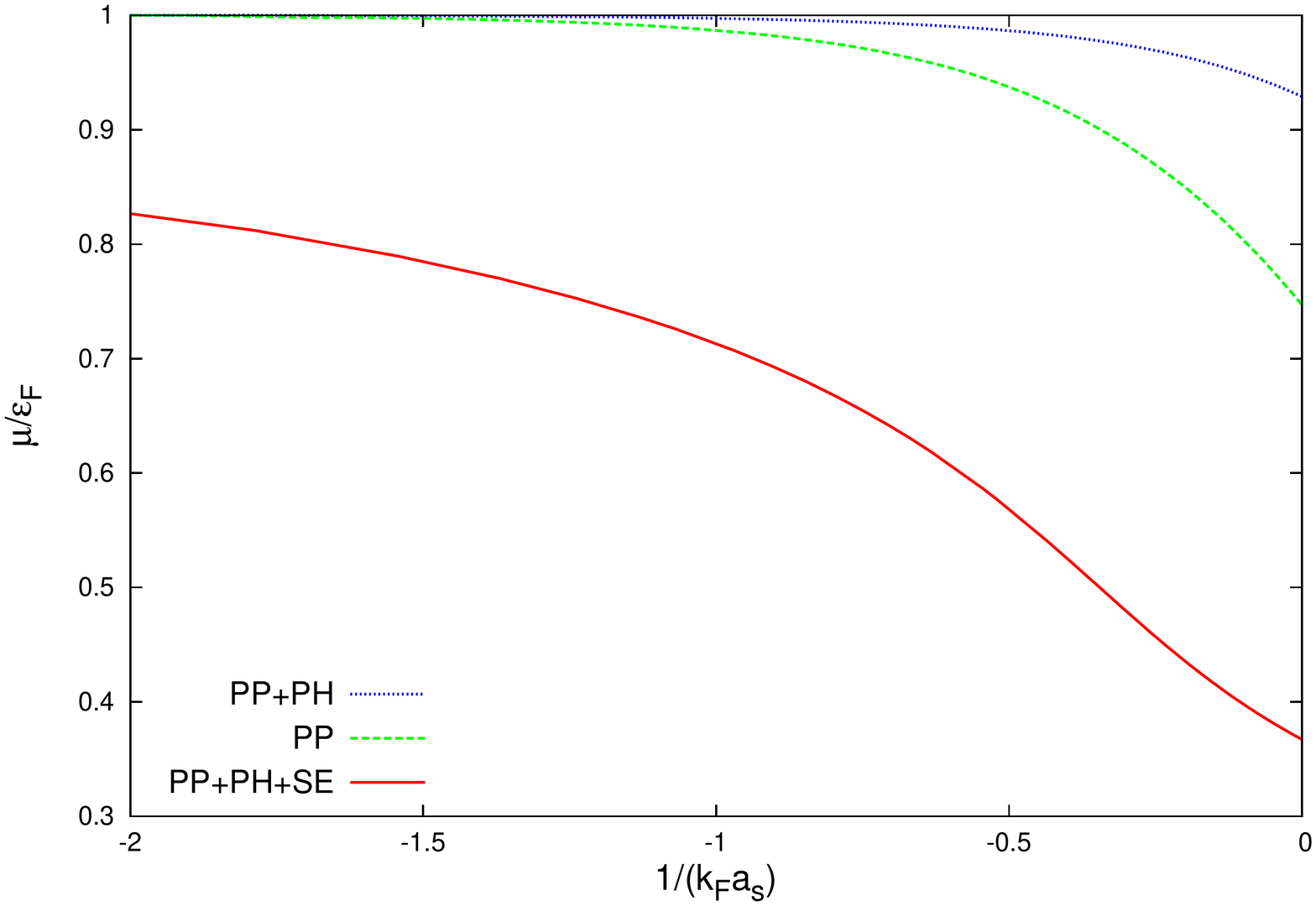}
\caption{$\mu(T_c)/\e_F$ as a function of the dimensionless scattering strength, $(k_F a_s)^{-1}$.
The meaning of each line is the same as in Fig.\ref{Fig5}.
}
\label{Fig6}
\end{figure}

   Flows of the self-energy are shown in  Fig. \ref{Fig4}  for inverse scattering lengths $(k_F a_s)^{-1}= -2, -1, 0$ at the corresponding critical temperatures. Figures \ref{Fig5} and \ref{Fig6} show our numerical results of the  critical temperature ($T_c/\varepsilon_F$) and the chemical potential ($\mu/\varepsilon_F$)  as a function of the inverse scattering length ($(k_F a_s)^{-1}<0$), respectively. 
    
For large $k$ $(\gtrsim k_F)$,  flow of the self-energy in Fig. \ref{Fig4} is well described by its asymptotic behavior given in (\ref{se_rpa02}).
 On the other hand,   flow of the self-energy almost stops for small $k$ $(\lesssim k_F)$. This is due to the fact that  approximate particle-hole symmetry for small $k$  protects the shift of the Fermi level. Fig. \ref{Fig4} shows that the magnitude of the self-energy $\sigma_0$ becomes larger as the coupling  becomes strong, but the saturation of $\sigma_k$ at $k\sim k_F$ is irrespective of the coupling strength.


As can be seen from Fig. \ref{Fig5},  PH correlation is the dominant source of the reduction of $T_c/\e_F$,  and its physical origin is the screening of the coupling strength.\footnote{Note that our numerical solutions of  $T_c/\e_F$ with the PP correlation and PP+PH correlation are smaller than the BCS formula, Eq. (\ref{BCSres}), and the GMB formula, Eq. (\ref{05}), respectively, valid {\it only} in the weak coupling limit.} 
We  note that this does not imply that the self-energy correction is negligible in the weak coupling region.
Since the flow of $\sigma_k$ stops for $k\lesssim k_F$,  most of its effect is absorbed  by shifting the Fermi level as $\mu\mapsto \mu_0=\mu+\sigma_0$.
The self-energy (SE) correction still leads to further reduction of  $T_c/\e_F$ since it makes $\e_F$ or the number density $n$ bigger as the coupling strength becomes stronger towards the unitary regime, but the effect is rather limited because $\sigma_0-\sigma_k$ for $k\sim k_F$ cannot be large due to the particle-hole symmetry, as discussed above. 

Fig. \ref{Fig6} shows that, the effect of the PH correlation on $\mu/\e_F$ is the opposite of what we find for $T_c/\e_F$. This is because the critical temperature $T_c$ of PP is higher than that of PP+PH for the same chemical potential $\mu$ due to the GMB correction. Therefore, the number of fermions in PP is larger than that of PP+PH. 
When the self-energy correction is taken into account in PP+PH+SE, the number density $n$ and hence the Fermi energy $\varepsilon_F$ increase for given $\mu$, so that a considerable decrease of $\mu/\e_F$ arises. It is interesting to see that even in the weak coupling limit, $\e_F$ is comparable with $\mu$, even though the self-energy has no significant effect on $T_c/\e_F$ in this regime.


\begin{table}[tq]
\centering
\begin{tabular}{c|cccp{0pt}|cccc}\hline
& \multicolumn{3}{c}{$\quad T_c/\e_F \quad$} & &  \multicolumn{3}{c}{$\quad \mu/\e_F\quad$}\\ 
\hline
$(k_F a_s)^{-1}$ & $-2$ &$-1$& $0$ && $-2$ &$-1$& $0$\\ \hline
PP       & 0.027  & 0.126 & 0.496  &&  1 & 0.987 & 0.747 \\
PP+PH &  0.012 & 0.058 & 0.276 &&   1 & 0.997 & 0.929  \\
PP+PH+SE&  0.012  & 0.053 & 0.237 & &  0.828 & 0.713 & 0.367 \\ \hline
\end{tabular}
\caption{\label{tab1}Numerical values of  $T_c/\e_F$ and $\mu/\e_F$ at $(k_F a_s)^{-1}=-2$, $-1$, and $0$ 
in different levels of approximation. The meaning of PP, PP+PH and PP+PH+SE are the same as those in Fig. \ref{Fig5}.} 
\end{table} 

In Table \ref{tab1} we pick up some values of
  $T_c/\varepsilon_F$ and $\mu/\varepsilon_F$ obtained from our numerical calculations at
  $(k_F a_s)^{-1}=-2$ (weak coupling), $(k_F a_s)^{-1}=-1$ (intermediate coupling)
   and at $(k_F a_s)^{-1}=0$ (unitarity).
 Although it is beyond our scope to predict a quantitatively correct $T_c$ in the unitary regime, 
 it is still instructive to compare our result at unitarity, $(T_c/\e_F,\mu/\e_F)=(0.237, 0.367)$,
 with the previous FRG results using auxiliary bosonic field,
 $(T_c/\e_F, \mu/\e_F)\simeq (0.264, 0.68)$ \cite{PhysRevB.78.174528}, $\simeq (0.276, 0.63)$
  \cite{ANDP:ANDP201010458}, $\simeq (0.248, 0.51)$ \cite{PhysRevA.81.063619,arXiv:1010.2890},
   and with the results of the $t$-matrix approach, 
     $T_c/\e_F \simeq 0.16$ \cite{PhysRevA.75.023610}, 
 and  $T_c/\e_F \simeq 0.217$ \cite{arXiv:1109.2307,PhysRevA.86.023610}.
 Note that quantum Monte Carlo  simulations\cite{PhysRevLett.96.160402,PhysRevLett.101.090402,PhysRevLett.103.210403,PhysRevA.82.053621,PhysRevB.76.165116} 
 for  $T_c/\e_F$ vary in the range  0.15$-$0.3.

\section{Summary and Concluding Remarks}\label{sec:conclusions}

We developed a fermionic FRG (f-FRG) method with a modified Litim's regulator 
to describe the superfluid phase transition for two-component fermions with a contact interaction. 
 By making vertex expansion of the 1PI effective action $\Gamma_k[\bar{\psi},\psi]$ up to the
 four-point vertex and solving the RG flow equation,
  we determined the critical temperature $T_c$ in the
   regime of negative scattering lengths using the Thouless criterion $[ \Gamma_{k=0}^{(4)}(p=0) ]^{-1}=0$.

In order to clarify the relation between  the FRG approach
and the conventional many-body theory such as the BCS theory + GMB correction,
 we have taken into account the particle-particle correlation, the particle-hole correlation and
 the self-energy correction step by step in the flow equations.
In agreement with the literature, we saw that the flow equation with a single PP correlation reduces to the BCS theory,
and we have also found that equipped with the PH correlation in the flow equation, a momentum-independent vertex
approximation precisely reproduces the GMB correction.  The  major part of the PH  contribution originated
   from  the region  $k \sim k_F$.

  To go beyond the BCS+GMB theory in our framework, we considered the flow of the constant part of the
 self-energy $\sigma_k$, together with the flow of the 4-point vertex  $\Gamma_{k}^{(4)}(p=0)$.
  We found that the self-energy decreases slowly as  $\sigma_k \sim 1/(k-3\pi/(8a_s))$ for $k\rightarrow \infty$, while
  it saturates as $\sigma_k \sim \sigma_0$ for  $k$ $(\lesssim k_F)$ due to approximate  particle-hole symmetry.
  We also carried out a numerical calculation with a hybrid approach, in which
 $\Gamma_{k}^{(4)}(p=0)$ is determined from the flow equation, while the
 $p$-dependent part of $\Gamma_{k}^{(4)}(p)$ necessary to reproduce  correct asymptotic behavior of $\sigma_k$  is evaluated using PP-RPA.  
Resultant value of  $T_c/\e_F$  does not receive  large correction from the self-energy except for the unitary regime ($1/(k_Fa_s) \rightarrow 0$).
 On the other hand, $\mu/\e_F$ shows a large reduction  by the self-energy correction due to the increase of the Fermi energy given $\mu$ even for relatively weak-coupling region $(k_F a_s)^{-1}\lesssim-1$. Extrapolation to unitarity gives $T_c/\e_F=0.237$ and $\mu/\e_F=0.367$.

  We note that the approximations employed in the present paper, particularly the assumption of constant
   self-energy, is valid only in  the  regime $1/(k_F a_s) < 0$. To enter into the BEC side
    ($1/(k_F a_s) > 0$), momentum-dependent  self-energy is crucial in f-FRG method, since
    it provides the information of composite bosons in the number equation.   Therefore, solving the flow equations with
    momentum-dependent self-energy  is one of the important problems to be studied in the future.
  More sophisticated treatments could also be employed, e.g. the use of the two-particle-irreducible (2PI) formalism \cite{PhysRevB.75.085102,PhysRevD.71.025010,Dupuis2005} combined with the f-FRG method is expected to be an efficient way to describe the superfluid phase transition, with its power of resumming an even wider class of Feynman diagrams.
Another direction to study the superfluid phase transition is to consider f-FRG with an IR regulator in the fermion vertex \cite{Tanizaki:2013yba}, which enables to treat the nontrivial momentum dependence of the fermion self-energy. 
These methods may open new ways for unbiased studies of the BCS-BEC crossover in the future by extending the f-FRG approach given in the present paper. 


\section*{Acknowledgements}
Y. T. is supported by JSPS Research Fellowships for Young Scientists. 
G. F. is supported by the Foreign Postdoctoral Research program of RIKEN. 
T. H. is partially supported by RIKEN iTHES project.
This work was partially supported  by the Program for Leading Graduate Schools, MEXT, Japan.
\appendix
\section{RPA results for the vertex function at finite $k$}
\renewcommand{\theequation}{A\arabic{equation}}
\setcounter{equation}{0}

In this appendix, we show the results of 
the four-point vertex function in particle-particle RPA (PP-RPA). 
According to (\ref{rpa03}), the basic equation is given by 
\bea
{1\over \Gamma_k^{(4,\mathrm{PP})}(p)}={1\over g}
+\int\limits_{\{|\bm{l}|<\Lambda_{\mathrm{UV}}\}} {\diff^3\bm{l}\over (2\pi)^3}
{1-\sum_{\pm}n_F\left(({\bm{p}\over 2}\pm\bm{l})^2-\mu+ R_k ( {\bm{p}\over 2}\pm\bm{l})\right)
\over 
ip^0+\sum_{\pm}\left\{({\bm{p}\over 2}\pm\bm{l})^2-\mu+R_k({\bm{p}\over 2}\pm\bm{l})\right\}}, 
\label{RPA01}
\eea
where we take $2m=1$ for simplicity, and 
$g$ is the bare coupling constant, as we indicated after (\ref{rpa03}). 
We consider the expansion given in (\ref{Eq:expan}) with the parameterization denoted as 
\begin{equation}
\Gamma_k^{(4,\mathrm{PP})}(p)=-{Z_k\over ip^0+S_k^{(1)} |\bm{p}|+S_k^{(2)} \bm{p}^2+|\mu^B_k|},
\label{RPA02}
\end{equation}
where parameters $Z_k$, $S_k^{(1)}$, $S_k^{(2)}$, and $|\mu^B_k|$ are positive.

Let us first consider $Z_k$, which may be regarded as  the wave function renormalization constant of Cooper pairs. 
According to (\ref{RPA02}), we obtain,
\begin{eqnarray}
Z_k^{-1}&=&-{\p \Gamma_k^{(4,\mathrm{PP})-1}\over \p (i p^0)}(0)=\int {\diff^3\bm{l}\over (2\pi)^3}
{1-2 n_F (\bm{l}^2-\mu+R_k(\bm{l})) \over [2(\bm{l}^2-\mu+R_k(\bm{l}))]^2}.
\label{RPA03}
\end{eqnarray}
Using the notation $k_{\pm}^2=\pm k^2+\mu$, we find that 
\begin{eqnarray}
Z_k^{-1}&=&{1\over 8\pi^2}\left[
{1\over 2}\left({k_+\over k_+^2-\mu}+{1\over \sqrt{\mu}}\tanh^{-1}{\sqrt{\mu}\over k_+}\right)
+{k_+^3-\sqrt{\mu}^3\over 3}{1-2n_F(k_+^2-\mu)\over (k_+^2 -\mu)^2}\right.\nonumber\\
&&\left.
-{1\over 2}\Re\left({k_-\over k_-^2-\mu}+{1\over \sqrt{\mu}}\tanh^{-1}{k_-\over \sqrt{\mu}}\right)
+{\sqrt{\mu}^3-\Re k_-^3\over 3}{1-2n_F(k_-^2-\mu)\over (k_-^2-\mu)^2}\right.\nonumber\\
&&\left.
-2\left(\int_0^{\Re k_-}+\int_{k_+}^{\infty}\right)l^2\diff l{n_F(l^2-\mu)\over (l^2-\mu)^2}
\right]. 
\label{RPA04}
\end{eqnarray}
According to the expression (\ref{RPA04}), asymptotic behavior of $Z_k$ in the large $k$ limit is 
given by 
\begin{equation}
Z_k^{-1} = {1\over 6\pi^2 k_+}\left(1+{\mu\over k_+^2}\right)+\mathcal{O}(1/k^4). 
\label{RPA05} 
\end{equation}
In order to get the spatial momentum dependence of the four-point function, 
we must perform the integration (\ref{RPA01}) with great care concerning the singularities associated with
the IR regulator. 
The coefficient of the $|\bm{p}|$-linear term is given by 
\begin{equation}
S_k^{(1)}={\mu Z_k \over 8\pi^2 }\left({\tanh{\beta\over 2}k^2 \over k^2}
-{\beta/2\over \cosh^2{\beta\over 2}k^2}\right). \label{RPA08}
\end{equation}
This term originates from the discontinuity of the regulator $R_k(\bm{l})$ at $|\bm{l}|=\sqrt{\mu}$, 
and it vanishes at $k=0$. 
As for the quadratic term in ${\bm p}$, we find that 
\begin{eqnarray}
S_k^{(2)}&=&{Z_k\over 16\pi^2 }\left[
{1\over 2}\left({k_+\over k_+^2-\mu}+{1\over \sqrt{\mu}}\tanh^{-1}{\sqrt{\mu}\over k_+}\right)
+{k_+^3\over 3}{1-2n_F(k_+^2-\mu)\over (k_+^2-\mu)^2} \right. \nonumber\\
&&\left. +{2\over 3}{k_+^3 n'_F(k_+^2-\mu)\over k_+^2-\mu}
-{1\over 2}\left({\Re k_-\over k_-^2-\mu}+
{1\over \sqrt{\mu}}\tanh^{-1}{\Re k_-\over \sqrt{\mu}}\right) \right.\nonumber\\
&&\left.+2\left(\int_{0}^{\Re k_-}+\int_{k_+}^{\infty}\right)l^2\diff l
\left(-{n_F(l^2-\mu)\over (l^2-\mu)^2}+{n'_F(l^2-\mu)+{2\over3}l^2n''_F(l^2-\mu)\over l^2-\mu}\right)
\right.\nonumber\\
&&\left. -{\Re k_-^3\over 3}
\left({2 n'_F(k_-^2-\mu)\over k_-^2-\mu}+{1-2n_F(k_-^2-\mu)\over (k_-^2-\mu)^2}\right) 
\right]. 
\label{RPA09}
\end{eqnarray}
The large-$k$ behavior of this quantity reads as 
\begin{equation}
S_k^{(2)}={1\over 2}+{\mu^{3/2}\over 4k^3}+\mathcal{O}(1/k^5),
\label{RPA11}
\end{equation}
which implies that the effective mass of the two-particle resonance is nothing but twice of the fermion mass, if $k$ is sufficiently large. 

Although we are not using the PP-RPA estimate of $|\mu^B_k|$ in the text, we show it here 
for the sake of completeness:
\be
|\mu^B_k|=Z_k \left[{1\over 8\pi a_s}+\int{\diff^3\bm{l}\over(2\pi)^3}\left({1-2n_F(\bm{l}^2-\mu+R_k(\bm{l}))\over 2(\bm{l}^2-\mu+R_k(\bm{l}))}-{1\over 2\bm{l}^2}\right)\right], 
\ee
whose  large-$k$ asymptotic behavior  is given by
\be
\mu^B_k={k^2}-\mu-{3\pi \over 4a_s}\sqrt{k^2+\mu}+\mathcal{O}(1/k). 
\ee

\section{Asymptotic behavior of the PP and PH contributions\label{app:asymp}}
\renewcommand{\theequation}{B\arabic{equation}}
\setcounter{equation}{0}

Here we list the analytic formulas for the asymptotic $k$ behavior of 
the right-hand side of  (\ref{gmb03}). 
Considering the $k$-derivative of the inverse coupling, $\p_k(1/\Gamma_k^{(4)})$, 
the first term in the right-hand side of (\ref{gmb03}), the PP contribution, reads
\begin{eqnarray}
&&{\p}_k\int_{l}^{(T)}{1\over [G^{-1}+R_k](l)[G^{-1}+R_k](-l)}\nonumber\\
&\simeq&
\left\{
{\renewcommand\arraystretch{2}
\begin{array}{cc}
\displaystyle-\left({m\over 3\pi^2}+{m\over 2\pi}{k_F^2\over k^2}\right),& (k^2/2m\gg k_F^2/2m,\pi T), \\
\displaystyle-{m\over \pi^2}{k_F\over k}, & (\pi T\ll k^2/2m \ll k_F^2/2km), \\
\displaystyle-{k_F\over 96 \pi^2 m^2 T^3}k^5,& (k^2/2m\ll k_F^2/2m,\pi T),
\end{array}}
\right.
\end{eqnarray}
with $k_F^2=2m\mu$.
The second term in the right-hand side of (\ref{gmb03}), the PH contribution, reads 
\begin{eqnarray}
&&{\p}_k
\int{\diff^2\hat{\bm{q}}\over 4\pi}{\diff^2\hat{\bm{q}}'\over 4\pi}
\int_{l}^{(T)}\hspace{0.2em}{1
\over [G^{-1}+R_k](l) [G^{-1}+R_k](q-q'+l)}\nonumber\\
&\simeq&
\left\{
{\renewcommand\arraystretch{2}
\begin{array}{cc}
\displaystyle {8 m\over 15\pi^2}{k_F^3\over k^3},& (k^2/2m\gg k_F^2/2m,\pi T), \\
\displaystyle {3m\over 4\pi^2}{k\over k_F}, & (\pi T\ll k^2/2m\ll k_F^2/2m), \\
\displaystyle {f(T/\mu)\over 2\pi^5 m k_F T^2}k^5,& (k^2/2m\ll k_F^2/2m,\pi T), 
\end{array}}
\right.
\end{eqnarray}
where $f(x)$ is given by 
\begin{eqnarray}
f(x)&=&
\sum_{n=1}^{\infty}{1\over (2n-1)^3}\left[
\int_0^2\diff Q \left(\pi-\tan^{-1}{(2n-1)\pi x \over 1-(Q-1)^2}
-\tan^{-1}{(2n-1)\pi x \over (Q+1)^2-1}\right)\right.\nonumber\\
&&\left.+{(2n-1)\pi x \over 16 }\ln\left(1+\left({ 8 \over (2n-1)\pi x}\right)^2\right)\right].
\end{eqnarray}

\end{document}